\documentclass[12pt]{article}
\usepackage{amsmath}
\usepackage{amsfonts}
\usepackage{multicol}
\usepackage{a4}

\begin{document}

\vspace*{-3cm}

\begin{center} 
  {\bf \Large Newton's Constant isn't constant\footnote{To appear in
  the Annual Report 2000 of the International School in Physics and
  Mathematics, Tbilisi, Georgia.}}\\[5mm]
  {\bf \normalsize Martin Reuter}\\[3mm]
  {\it \normalsize University of Mainz, Germany}
\end{center} 

\vspace{5mm}

According to Newton's law of gravity, two masses interact with each
other via a central force which can be derived from the potential
$V_{\rm N}\left(r \right)= - G m_{1}m_{2}/r$ where $G$ is a universal
constant known as Newton's constant. Likewise, according to Coulomb's
law, two electrically charged particles with charges $n_{1}e$ and
$n_{2}e$ interact via the central potential $V_{\rm Cb}\left(r
\right)=\alpha n_{1}n_{2}/r$ where $e$ is the elementary charge and
where $ \alpha \equiv e^{2}/4\pi $ is nowadays referred to as the
fine structure constant. Thus, in classical physics, the gravitational
and the electrostatic interactions are described by exactly the same
$1/r$-law, and their respective coupling strengths are determined by
the two universal constants $G$ and $\alpha$ which enter the equations
in an analogous fashion.

However, from the point of view of modern quantum field theory we know
that $e$ and $\alpha$ are not really constants but are more
appropriately considered scale dependent or ``running" quantities. In
quantum electrodynamics (QED) the charge $e$ of a positron is a
function $e\left(k \right)$ depending on the ``renormalization scale"
$k$, a parameter with the dimension of a mass which specifies the
resolution of the ``microscope" with which we probe the system. The
physical mechanism behind the scale dependence of the electric charge
is easy to understand. The combination of Quantum Mechanics and
Special Relativity converts the vacuum of electrodynamics into a sea
of virtual electron-positron pairs which are con\-ti\-nu\-ous\-ly
created and annihilated. When we immerse an external test charge into
this sea it gets polarized in very much the same way as an ordinary
dielectric. The polarization cloud of the virtual $e^{+}/e^{-}$-pairs
surrounding the test charge tends to screen it, and it appears to be
larger at small distances and smaller at large distances. In an
experiment which resolves length scales $ \ell \equiv k^{-1}$ one
measures the effective charge $ e\left(k \right)$ which includes the
effect of this polarization of the vacuum.

As a consequence of the same screening mechanism the classical Coulomb
potential is replaced by a more complicated quantum corrected
potential, the Uehling potential $ V_{\rm Uehling}\left(r \right)$. At
least in the limit of massless electrons, this potential is directly
related to the running charge. Considering an electron in the field of
a positron, say, one starts from the classical potential energy $
V_{\rm Cb}\left(r \right)=- e^{2}/4\pi r$ and replaces $ e^{2}$ by the
running gauge coupling in the one-loop approximation:
\begin{displaymath}
e^{2}\!\left(k \right) = 
e^{2}\!\left(k_{0} \right)\left[1-b\, {\ln}\left(k/k_{0} \right)
\right]^{-1}, 
\qquad b\equiv e^{2}\!\left(k_{0} \right)/6\pi ^{2}. 
\end{displaymath}
(We are using units such that $ \hbar =c=1$.)  The crucial step is to
identify the renormalization point $k$ with the inverse of the
distance $r$. This is possible because in the massless theory $r$ is
the only dimensionful quantity which could define a scale. The result
of this substitution reads
\begin{displaymath}
V_{\rm Uehling}\left(r \right) = -e^{2}\!\left(r^{-1}_{0}
\right)\left[1+b\, {\ln}\left(r_{0}/r \right)+O\left(e^{4}
  \right)\right]/4\pi r 
\end{displaymath}
where the IR reference scale $ r_{0}\equiv 1/k_{0}$ has to be kept
finite in the massless theory. Our result is the correct (one-loop,
massless) Uehling potential which is usually derived from the
polarization tensor of the photon. Obviously the position dependent
renormalization group improvement $e^{2}\rightarrow e^{2}\left(k
\right),\, k\propto 1/r$ encapsulates the most important effects the
quantum fluctuations have on the electric field produced by a point
charge.

Because of the analogy between $\alpha$ and $G$ it is natural to ask
if there are similar quantum effects which render Newton's constant
scale dependent. Clearly the first step towards an answer to this
question consists of replacing Newtonian gravity by General
Relativity. Here the relevant field-source relation is Einstein's
equation
\begin{displaymath}
R_{\mu \nu }-\frac{1}{2}\, g_{\mu \nu }\, R = 8\pi G \, T_{\mu \nu }
\end{displaymath}
which reproduces Newton's law in an appropriate limit. In General
Relativity, too, $G$ is a universal constant, the coupling constant of
the gravitational self-interaction and of the gravity-matter
interaction.

Contrary to the situation in electrodynamics we have no consistent
fundamental quantum field theory of gravity at our disposal yet.
Nevertheless, guided by the analogy with the running electric charge,
it is tempting to speculate on how quantum gravitational effects might
modify Newton's law and lead to a scale dependence of $G$. It is
plausible to assume that in the large distance limit the leading
quantum effects are described by quantizing the linear fluctuations of
the metric, $g_{\mu \nu}$. One obtains a free field theory in a
possibly curved background spacetime whose elementary quanta, the
gravitons, carry energy and momentum. The vacuum of this theory will
be populated by virtual graviton pairs, and the central question is
how these virtual gravitons respond to the perturbation by an external
test body which we immerse in the vacuum. Assuming that also in this
situation gravity is universally attractive, the gravitons will be
attracted towards the test body. Hence it will become ``dressed" by a
cloud of virtual gravitons surrounding it so that its effective mass
seen by a distant observer is larger than it would be in absence of
any quantum effects. This means that while in QED the quantum
fluctuations {\it screen} external charges, in quantum gravity they
have an {\it antiscreening} effect on external test masses. This
entails Newton's constant becoming a scale dependent quantity $
G\!\left(k \right)$ which is small at small distances $ \ell \equiv
k^{-1}$, and which becomes large at larger distances. This behavior is
similar to the running of the nonabelian gauge coupling in Yang-Mills
Theory.

Can we verify these heuristic arguments within a consistent theory? In
many of the traditional approaches to quantum gravity the
Einstein-Hilbert term $ \int d^{4}x \sqrt{-g} R$ has been regarded as
a fundamental action which should be quantized along the same lines as
the familiar renormalizable field theories in flat space, such as QED
for example. It was soon realized that this program is not only
technically rather involved but also leads to severe conceptual
difficulties. In particular, the nonrenormalizability of the theory
hampers a meaningful perturbative analysis. While this does not rule
out the possibility that the theory exists nonperturbatively, not much
is known in this direction. However, it could also be argued that
gravity, as we know it, should not be quantized at all, because
Einstein gravity is an effective theory which results from quantizing
some yet unknown fundamental theory. If so, the Einstein-Hilbert term
is an effective action analogous to the Heisenberg-Euler action in
QED, say, and it should not be compared to the ``microscopic" action
of electrodynamics.

It seems not unreasonable to assume that the truth lies somewhere
between those two extreme points of view, i.e., that Einstein gravity
is an effective theory which is valid near a certain nonzero momentum
scale $k$. This means that it arises from the fundamental theory by a
``partial quantization" in which only excitations with momenta larger
than $k$ are integrated out, while those with momenta smaller than $k$
are not included. (The interpretation of the Einstein-Hilbert term as
a fundamental or an ordinary effective action is recovered in the
limits $ k \rightarrow \infty $ and $k \rightarrow 0 $, respectively.)
By definition, an ``effective theory at scale $k$", when evaluated at
tree level, should correctly describe all gravitational phenomena
which involve a typical momentum scale $k$ acting as a physical
infrared cutoff. Only if one is interested in processes with momenta
$k' \ll k$, loop calculations become necessary; they amount to
integrating out the missing field modes in the momentum interval $
\left[k', k \right]$.

In ref.\cite{mr} it was proposed to regard the scale-dependent action
for gravity, henceforth denoted $ \Gamma _{k}\left[g_{\mu \nu }
\right]$ (``effective average action"), as a Wilsonian effective
action which is obtained from the fundamental (``microscopic") action
$S$ by a kind of coarse-graining analogous to the iterated block-spin
transformations which are familiar from lattice systems. In the
continuum, $ \Gamma _{k }$ is defined in terms of a modified Euclidean
functional integral over $ e^{-S}$ in which the contributions of all
field modes with momenta smaller than $k$ are suppressed. In this
manner $ \Gamma _{k }$ interpolates between $ S$ (for $k \rightarrow
\infty$) and the standard effective action $\Gamma$ (for $k
\rightarrow 0$). The trajectory in the space of all action functionals
can be obtained as the solution of a certain functional evolution
equation, the exact renormalization group (RG) equation. Its form is
independent of the action $S$ under consideration. The latter enters
via the initial conditions for the renormalization group trajectory;
it is specified at some UV cutoff scale $ \Lambda : \Gamma _{\Lambda
}=S$. If $S$ is a {\it fundamental} action, $\Lambda$ is sent to
infinity at the end. The renormalization group equation can also be
used to evolve {\it effective} actions, known at some point $\Lambda$,
towards smaller scales $ k < \Lambda $. In this case $\Lambda$ is a
fixed, finite scale. In this framework, the (non)renormalizability of
a theory is seen as a global property of the renor\-ma\-li\-za\-tion
group flow for $ \Lambda \rightarrow \infty $. The evolution equation
by itself is perfectly finite and well behaved in either case because
it describes only infinitesimal changes of the cutoff.

In the construction of ref.\cite{mr} the modified functional integral
over $ e^{-S}$ is similar to the standard gauge-fixed path-integral of
Euclidean gravity in the background gauge. The crucial new ingredient
is a built-in infrared (IR) cutoff which suppresses the contributions
from long-wavelength field modes. It is implemented by giving a
$k$-dependent and mode-dependent mass $ {\cal R}_{k}\left(p^{2}
\right)$ to the modes with covariant momentum $p$. Inside loops, it
suppresses the small-$p$ contributions. The function ${\cal
R}_{k}\left(p^{2} \right)$ has to satisfy ${\cal R}_{k}\left(p^{2}
\right) \rightarrow 0$ for $k \rightarrow 0$ and ${\cal
R}_{k}\left(p^{2} \right) \propto k^{2}$ for $k \gg p$, but is
arbitrary otherwise. (In practice the exponential cutoff ${\cal
R}_{k}\left(p^{2} \right) \propto p^{2} [\exp\left(p^{2}/k^{2}
\right)-1]^{-1}$ is convenient.)

In order to obtain a functional $ \Gamma _{k}\left[g \right]$ which is
invariant under general coordinate transformations the background
gauge formalism is employed. This means that we actually RG-evolve an
action $ \Gamma _{k}\left[g, \bar g \right] $ which depends on both
the ``ordinary" metric $ g_{\mu \nu }$ and on a background metric
$\bar g_{\mu \nu}$. The standard action is recovered by setting $\bar
g = g$, i.e.  $ \Gamma _{k}\left[g \right] \equiv \Gamma
_{k}\left[g,g\right]$. The exact renormalization group equation for
$\Gamma _{k}\left[g, \bar g\right]$ reads (see ref.\cite{mr} for
details):
\begin{displaymath} 
\partial _{t}\Gamma _{k}\left[g, \bar g \right] =   \frac{1}{2}{\rm
Tr} 
\left[\left(\kappa^{-2}\Gamma _{k}^{\left(2 \right)}\left[g, \bar g
    \right]+     
{\cal R}_{k}^{\rm grav}\left[\bar g \right] \right)^{-1} \partial
_{t}{\cal R}_{k}^{\rm grav} 
\left[\bar g \right] \right] 
\end{displaymath}
\begin{displaymath}
- {\rm Tr} \left[\left(- {\cal M}\left[g, \bar g \right] + {\cal
    R}_{k}^{\rm gh} 
\left[\bar g \right]\right)^{-1} \partial _{t}{\cal R}_{k}^{\rm
gh}\left[\bar g \right]\right]  
\end{displaymath}
with the ``renormalization group time" $t\equiv {\rm ln} k$. Here $
\Gamma ^{\left(2 \right)}_{k }$ stands for the Hessian of $\Gamma_{k}$
with respect to $ g_{\mu \nu }$ at fixed $\bar g_{\mu \nu }$, and
$\cal M$ is the Faddeev-Popov ghost operator. The operators ${\cal
R}^{\rm grav}_{k }$ and ${\cal R}^{\rm gh}_{k }$ implement the IR
cutoff in the graviton and the ghost sector. They obtain from $ {\cal
R}_{k} \left(p^{2} \right)$ by replacing the momentum square $p^{2}$
with the graviton and ghost kinetic operator, respectively.

Nonperturbative solutions to the above RG-equation (which do not
require any expansion in $G$) can be obtained by the method of
``truncations". This means that one projects the RG flow $k \mapsto
\Gamma_{k}$ in the infinite dimensional space $ \{\Gamma \left[ \cdot
\right] \}$ of all action functionals onto some finite dimensional
subspace which is particularly relevant. In this manner the functional
RG-equation becomes an ordinary diffe\-ren\-tial equation for a finite
set of generalized couplings which serve as coordinates on this
subspace. In ref.\cite{mr} we projected on the 2-dimensional subspace
spanned by the operators $ \sqrt{g}$ and $ \sqrt{g}R$
(``Einstein-Hilbert truncation"). This truncation of the ``theory
space" amounts to considering only actions of the form
\begin{displaymath}
\Gamma _{k} \left[g, \bar g \right] = \left(16\pi G\left(k \right)
\right)^{-1} 
\int\limits_{}^{}d^{4}x\sqrt{g}  \{-R\left(g \right)+2 \bar \lambda
\left(k \right) \} + 
S_{\rm gf}\left[g, \bar g \right]
\end{displaymath}
where $ G\left(k \right)$ and $ \bar \lambda \left(k \right)$ denote
the running Newton constant and cosmological constant, respectively,
and where $ S_{\rm gf}$ is the classical background gauge fixing term.
More general (and, therefore, more precise) truncations would include
higher powers of the curvature tensor as well as nonlocal terms, for
instance.

In ref.\cite{mr} we inserted the Einstein-Hilbert ansatz into the
RG-equation and derived the coupled system of equations for $ G\left(k
\right)$ and $ \bar \lambda\left(k \right)$. It is rather complicated
and we shall not write it down here. If $ \bar \lambda \ll k^{2}$ for
all scales of interest, it simplifies considerably and boils down to a
simple equation for the dimensionless Newton constant $ g\left(k
\right) \equiv k^{2} G\left(k \right)$ \cite{bh}:
\begin{displaymath}
\frac{d}{dt} g\left(t \right) = \beta \left(g\left(t \right) \right) ,
\quad  
\beta \left(g \right) = 2g  \frac{1-\omega^{\prime} g} {1-B_{2}g}
\end{displaymath}
For the exponential cutoff, the constants entering the beta-function
are \cite{bh}:
\begin{displaymath}
\omega ^{\prime} \equiv \omega + B_{2}, \quad 
\omega = \frac{4}{\pi } \left(1 - \frac{\pi ^{2}}{144} \right) , 
\quad
B_{2} = \frac{2}{3\pi}
\end{displaymath}
The above evolution equation for $g$ displays two fixed points $
g_{\ast } , \beta \left(g_{\ast } \right) = 0$. There exists an
infrared attractive (gaussian) fixed point at $ g^{\rm IR}_{\ast} = 0$
and an ultraviolet attractive (nongaussian) fixed point at
\begin{displaymath}
g^{\rm UV}_{\ast} = \frac{1}{\omega ^{\prime}}
\end{displaymath}
This latter fixed point is a higher dimensional analog of the Weinberg
fixed point \cite{wein} known from $ \left(2 + \epsilon
\right)$-dimensional gravity.

The UV fixed point separates a weak coupling regime $ \left(g< g^{\rm
  UV} _{\ast}\right)$ from a strong coupling regime where $ g> g^{\rm
UV} _{\ast}$. Since the $\beta$-function is positive for $ g \in
\left[0, g^{\rm UV}_{\ast} \right]$ and negative otherwise, the
renormalization group trajectories $ k \mapsto g\left(k \right)$ fall
into the following three classes:
\begin{itemize}
\item[(i)] Trajectories with $ g\left(k \right) < 0$ for all $k$. They
  are attracted towards $ g^{\rm IR}_{\ast}$ for $ k \rightarrow 0$.
\item[(ii)] Trajectories with $ g\left(k \right) >g^{\rm UV}_{\ast}$
  for all $k$. They are attracted towards $ g^{\rm UV}_{\ast}$ for $k
  \rightarrow \infty$.
\item[(iii)] Trajectories with $ g\left(k \right) \in \left[0, g^{\rm
    UV}_{\ast} \right]$ for all $k$. They are attracted towards $
  g^{\rm IR}_{\ast} = 0$ for $k \rightarrow 0$ and towards $ g^{\rm
  UV}_{\ast}$ for $ k \rightarrow \infty$.
\end{itemize}
Only the trajectories of type (iii) are relevant for us. We shall not
allow for a negative Newton constant, and we also discard solutions of
type (ii). They are in the strong coupling region and do not connect
to a perturbative large distance regime. (See ref.\cite{so} for a
detailed numerical investigation of the phase diagram.)

The trajectories of type (iii) cannot be written down in closed form
but, returning to the dimensionful quantity $G$, a numerically rather
precise approximation is given by
\begin{displaymath}
G\!\left(k  \right) = \frac{G\!\left(k _{0} \right)}{1+\omega \,
G\!\left(k _{0} \right) \left[k ^{2} - k ^{2} _{0}\right]} 
\end{displaymath}
We shall set $k_{0} = 0$ for the reference scale. At least within the
Einstein-Hilbert truncation, $ G\!\left(k \right)$ does not run any
more between scales where the Newton cons\-tant was determined
experimentally (laboratory scale, scale of the solar system, etc.) and
$k \approx 0$ (cosmological scale). Therefore we can identify $ G_{0}
\equiv G\left(k _{0} = 0 \right)$ with the experimentally observed
value of the Newton constant. From
\begin{displaymath}
G\left(k  \right) = \frac{G_{0}}{1 + \omega G_{0} k^{2}}
\end{displaymath}
we see that when we go to higher momentum scales $k$, $G\!\left(k
\right)$ decreases mono\-to\-ni\-cally. For small $k$ we have
\begin{displaymath}
G\left(k  \right) = G_{0} - \omega \,G^{2}_{0}  k^{2} + O\left(k^{4}
\right) 
\end{displaymath}
while for $ k^{2}\gg G^{-1}_{0}$ the fixed point behavior sets in and
$G \left(k \right)$ ``forgets" its infrared value:
\begin{displaymath}
G\left(k \right) \approx \frac{1}{\omega k^{2}}
\end{displaymath}
In ref.\cite{pol}, Polyakov had conjectured an asymptotic running of
precisely this form. If the UV fixed point can be confirmed by more
general truncations it means that Einstein gravity in 4 dimensions is
``asymptotically safe" in Weinberg's sense \cite{wein}.

Thus the heuristic arguments in favor of the antiscreening character
of pure quantum gravity seem to be correct \cite{gen}, and we may now
use our result for $ G\!\left(k \right)$ in order to RG-improve
Newton's potential. The leading large distance correction of $ V_{\rm
N}\left(r \right)$ is obtained by using the small-$k$ approximation
for $ G\!\left(k \right)$ and by setting $ k = \xi/r$, $\xi$ = const,
because $1/r$ is the only relevant IR cutoff if spacetime is
approximately flat. Reinstating factors of $ \hbar $ and $c$ for a
moment we find
\begin{displaymath}
V_{\rm imp}\left(r \right) = - G_{0}\frac{m_{1}m_{2}}{r} 
\left[1-\tilde\omega \frac{G_{0}\hbar }{r^{2}c^{3}}  +\ldots\right] 
\end{displaymath}
The constant $ \tilde \omega \equiv \omega \xi ^{2}$ is predicted to
be positive, but its precise value cannot be inferred from
RG-arguments alone. However, it was pointed out by Donoghue \cite{don}
that the standard perturbative quantization of Einstein gravity leads
to a well-defined, finite prediction for the leading large distance
correction to Newton's potential. His result reads
\begin{displaymath}
V\!\left(r \right) = - G_{0}\frac{m_{1}m_{2}}{r} 
\left[1 - \, \frac{G_{0}\left(m_{1} + m_{2} \right)}{2c^{2}r}  -
\hat \omega \, \frac{G_{0}\hbar }{r^{2}c^{3}} + \ldots \right]
\end{displaymath}
where $\hat \omega = 118/15 \pi$. The correction proportional to $
\left(m_{1} + m_{2} \right)/r $ is a purely kinematic effect of
classical general relativity, while the quantum correction $\propto \,
\hbar$ has precisely the structure we have predicted on the basis of
the renormalization group. Comparing the two potentials allows us to
determine the coefficient $\tilde \omega$ by identifying $\tilde
\omega = \hat \omega$.

With its only undetermined parameter fixed by Donoghue's asymptotic
calculation, we can now use our formula for $ G\left(r \right) \equiv
G\left(k\left(r \right) \right)$ in order to investigate gravity at
very short distances comparable to the Planck length. In refs.\ \cite
{bh, bhmass} the impact of the running of $G$ on the structure of
black holes has been considered as an example. In ref.\cite{bh} we
constructed a quantum-Schwarzschild black hole by improving $ G_{0}
\rightarrow G\!\left(r \right)$ in the classical Schwarzschild metric.
(In this context the correct identification of $k$ as a function of
$r$ is more subtle; a careful analysis yields $ k \propto r^{-1}$ for
$r \rightarrow \infty$, but $ k \propto r^{-3/2}$ for $r \rightarrow
0$.) The main features of the RG improved spacetime are as follows.

As far as the structure of horizons is concerned, the quantum effects
are small for very heavy black holes ($ M \gg m_{\rm Pl}$). They have
an event horizon at a radius $r_{+}$ which is close to, but always
smaller than the Schwarzschild radius $2G_{0}M$. Decreasing the mass
of the black hole the event horizon shrinks. There is also an inner
(Cauchy) horizon whose radius $r_{-}$ increases as $M$ decreases. When
$M$ equals a certain critical mass $M_{\rm cr}$ which is of the order
of the Planck mass the two horizons coincide. The near-horizon
geometry of this extremal black hole is that of $ AdS_{2}\times
S^{2}$. For $ M < M_{\rm cr}$ the spacetime has no horizon at all.

While the exact fate of the singularity at $r=0$ cannot be decided
within our present approach, it can be argued that either it is not
present at all or it is at least much weaker than its classical
counterpart. In the first case the quantum spacetime has a smooth de
Sitter core so that we are in accord with the cosmic censorship
hypothesis even if $ M < M_{\rm cr}$.

The conformal structure of the quantum black hole is very similar to
that of the classical Reissner-Nordstr\"om spacetime. In particular
its $ \left(r = 0 \right)$-hypersurface is timelike, in
contradistinction to the Schwarzschild case where it is spacelike.

The Hawking temperature $T_{\rm BH}$ of very heavy quantum black holes
is given by the semiclassical $1/M$-law. As $M$ decreases, $T_{\rm
BH}$ reaches a maximum at $ \widetilde M_{\rm cr} \approx 1.27 M_{\rm
cr}$ and then drops to $T_{\rm BH} = 0$ at $M = M_{\rm cr}$. The
specific heat capacity has a singularity at $\widetilde M_{\rm cr}$.
It is negative for $M > \widetilde M_{\rm cr}$, but positive for
$\widetilde M_{\rm cr} > M > M_{\rm cr}$. We argued that the vanishing
temperature of the extremal black hole leads to a termination of the
evaporation process once the black hole has reduced its mass to $M =
M_{\rm cr}$. This supports the idea of a cold, Planck size remnant as
the final state of the black hole evaporation.

For $M > M_{\rm cr}$, the entropy of the quantum black hole is a well
defined, monotonically increasing function of the mass. For heavy
black holes we recover the classical expression ${\cal A} / 4 G_{0}$.
The leading quantum corrections are proportional to $\ln\left(M/M_{\rm
  cr} \right)$.

\noindent
Acknowledgement: It is a pleasure to thank the organizers of the ISPM
conference for their cordial hospitality at Tbilisi and for
introducing us not only to the most recent developments in
mathematical physics but also to two thousand years of Georgian
history and culture.


\begin{thebibliography}{99}
  
\bibitem{mr} M. Reuter, Phys. Rev.  D57, 971 (1998) and hep-th/9605030
  
\bibitem{bh} A.\ Bonanno, M.\ Reuter, Phys.\ Rev.\ D62, 043008 (2000)
  and hep-th/0002196
  
\bibitem{wein} S.\ Weinberg, in {\it General Relativity, an Einstein
  Centenary Survey}, S. W. Hawking, W. Israel (Eds.), Cambridge Univ.
  Press, 1979
  
\bibitem{so} W. Souma, Prog. Theor. Phys. 102, 181 (1999)
  
\bibitem{pol} A. Polyakov, in {\it Gravitation and Quantization}, J.
  Zinn-Justin, B. Julia \\ (Eds.), North-Holland, 1995
  
\bibitem{don} J. F. Donoghue, Phys. Rev. Lett. 72, 2996 (1994); Phys.
  Rev. D50, 3874 (1994); H. W. Hamber, S. Liu, Phys. Lett. B357, 51
  (1995)
  
\bibitem{bhmass} A. Bonanno, M. Reuter, Phys. Rev. D60, 084011 (1999)
  and gr-qc/9811026
  
\bibitem{gen} For generalizations and the inclusion of matter fields
  see:\\
  D. Dou, R. Percacci, Class. Quant. Grav. 15, 3449 (1998); L. N.
  Granda,\\ S. D. Odintsov, Grav. Cosmol. 4, 85 (1998) and Phys. Lett.
  B409, 206 (1997);\\ A. Bytsenko, L. N. Granda, S. D. Odintsov, JETP
  Lett. 65, 600 (1997);\\ S. Falkenberg, S. D. Odintsov, Int. J. Mod.
  Phys. A13, 607 (1998)

\end{thebibliography}
\end{document}